\documentclass[aps,prl,reprint,twocolumn]{revtex4-2}

\usepackage{mathrsfs}  
\usepackage{amssymb}
\usepackage{color} 
\usepackage{ulem}
\usepackage{mathptmx}
\usepackage{enumitem}
\usepackage{amsmath}    
\usepackage{graphicx}   
\usepackage{verbatim}   
\usepackage{hyperref}   
\begin{document}
\title{Gravito-capillary pinning of pendant droplets under wet uneven surfaces}
\author{Etienne Jambon-Puillet}
\affiliation{LadHyX, CNRS, Ecole Polytechnique, Institut Polytechnique de Paris, Palaiseau, France}

\date{\today}

\begin{abstract}
Pendant drops spontaneously appear on the underside of wet surfaces through the Rayleigh-Taylor instability. These droplets have no contact line, they are connected to a thin liquid film with which they exchange liquid and are thus mobile: any perturbation will set them in motion. Here, using experiments, numerical simulations, and theory I show that pendant drops sliding under a slightly tilted wet substrate can pin on topographic defects, despite their lack of contact line. Instead, this pinning force has a gravito-capillary origin: liquid has to moves up or down and the interface has to deforms for the drop the pass the defect. I propose a semi-analytical model for arbitrary substrate topographies that matches the pinning force observed experimentally and numerically, without any fitting parameter. I finally demonstrate how to harness this pinning force to guide pendant drops on complex paths.
\end{abstract}

\maketitle

\begin{figure}[b]
    \begin{center}
        \includegraphics[width=\columnwidth]{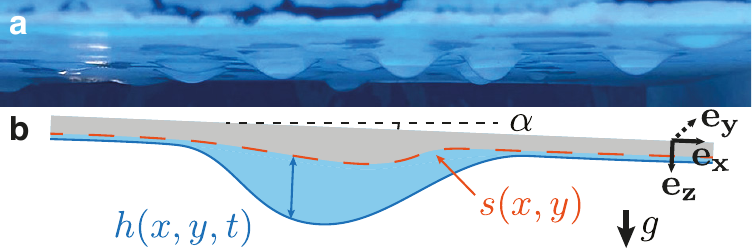}
        \caption{\textbf{a} Several pendant droplets on the ceiling of a swimming pool skimmer.
        \textbf{b} Schematic of a thin liquid film of thickness $h(x,y,t)$ that contains a pendant drop sliding under a surface with topography $s(x,y)$, inclined by an angle $\alpha$.
        } 
        \label{fg:fig1}
    \end{center}
\end{figure}

A thin liquid film covering the underside of a surface will spontaneously destabilize and form an array of pendant droplets~\cite{Yiantsios:1989,Fermigier:1992}. This process which can be easily observed in kitchens, bathrooms and other everyday life situations (e.g. Fig.~\ref{fg:fig1}a) has important consequences for many applications; it can impair the quality of coatings \cite{Weinstein:2004} or conversely be harnessed to pattern surfaces \cite{Marthelot:2018,Jambon:2021a}, it impacts geomorphological processes \cite{Ribe:2007,Dutta:2016,Ledda:2021}, and can be detrimental to engineering constructs with wet surfaces~\cite{Kaita:2010,vanEden:2017}. For these reasons, the Rayleigh-Taylor instability in thin viscous films has been thoroughly investigated~\cite{Yiantsios:1989,Fermigier:1992,Limat:1992,Yoshikawa:2019,Lerisson:2020} and various approaches to avoid it have been developed and rationalized with linear stability analysis~\cite{Burgess:2001,Lapuerta:2001,Alexeev:2007,Cimpeanu:2014,Trinh:2014,Brun:2015,Balestra:2018}. 

The late-time, non-linear dynamics of the fully formed pendant droplets has received less attention~\cite{Lister:2010,Jambon:2021b}. A distinctive feature of these drops is their absence of contact line, they are connected to a thinner macroscopic film with which they continuously exchange liquid. This makes them very mobile and under smooth, and uniformly wet substrates any tilt or perturbation will set them in motion and alter their growth rate~\cite{Lister:2010,Jambon:2021b}. However, surfaces in most practical situations are uneven. For instance, the ceiling of caves is covered by speleotherms~\cite{Ledda:2021}, and patterned surfaces have been used to control the local thickness of coatings~\cite{Marthelot:2018}. For sessile drops, surface roughness usually hinders their motion by pinning their contact line~\cite{Joanny:1984,Nadkarni:1992,Kalinin:2009,Park:2017}. Since pendant drops on the underside of wet surfaces lack a contact line, the effect of topographical defects on their motion is unclear.

In this letter, using experiments, numerical simulations, and theory, I show that topographic defects can pin a single pendant drop sliding under a slightly tilted uniformly wet substrate. The defect generates a pinning force which stops the drop if the inclination angle is too low. This critical inclination angle depends on the defect dimensions, and not on the drop size or underlying film thickness. I derive a gravito-capillary model that yields a semi-analytical prediction for the pinning force and critical angle that matches experiments and simulations without any fitting parameters. Locally, as the drop slides on the defect, fluid is displaced vertically and the interface is distorted creating potential energy variations responsible for the force. Finally, I demonstrate how to use this pinning force to guide the motion of pendant drops using the substrate topography.

\begin{figure*}[t]
    \begin{center}
        \includegraphics[width=\textwidth]{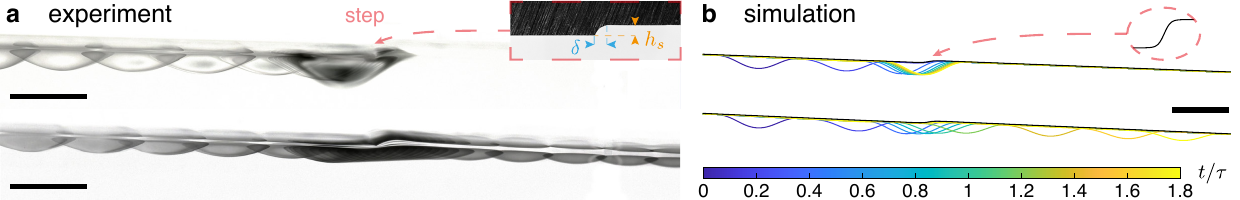}
        \caption{\textbf{a.} Side view chronophotography of two experiments showing a pendant drop sliding under a surface with a step of height $h_s=330$ $\mu$m and sharpness $\delta\approx 400$ $\mu$m (see inset), pre-wet with a film of thickness $h_0\approx 60$ $\mu$m, at two different inclination angles $\alpha=\{2.2,2.5\}$ deg. The images include the drop reflection on the wet substrate. Scale bar $5$ mm, time interval $2$ min. 
        \textbf{b.} Numerical replication of the experiment in a (identical parameters except for the step sharpness $\delta=\ell_c$). The profiles are slices through the center of the 3d simulations, color codes the time, scale bar $5$ $\ell_c$. The inset shows the step dilated vertically by a factor $5$. See also movie S1 and S2~\cite{Note1}.
        } 
        \label{fg:fig2}
    \end{center}
\end{figure*}

The problem considered here is schematized in Fig.~\ref{fg:fig1}b. The underside of a solid with a non-flat topography, slightly tilted by an angle $\alpha$, is wet by a thin-layer of liquid over which lies a pendant drop. Experimentally, I create controlled surface defects by milling PMMA plates and use silicone oil (density $\rho=971$ kg/m$^3$, surface tension $\gamma=20.5$ mN/m, viscosity $\eta=1$ Pa.s) as a fluid. Pre-wetting is achieved by spin-coating oil on the substrate which results in a uniform film thickness $h_0$, except close to the defects~\cite{Stillwagon:1990,Gaskell:2004}. The substrate is then mounted upside down to a rotation stage. A single pendant droplet of amplitude  $A_0\sim \ell_c$ is generated by adding oil on the spin-coated film with a syringe. Here $\ell_c=\sqrt{\gamma/(\rho g)}\approx 1.47$ mm denotes the capillary length with $g$ the gravitational acceleration. The substrate is finally tilted by an angle $\alpha$ and the drop dynamics is recorded (see Supplemental Material for experimental methods~\footnote{See Supplemental Material for experimental, numerical, and theoretical methods, additional results and movies of the experiments and simulations.}). Note that the spin-coated film is sufficiently thin and smooth to neglect the growth of the Rayleigh-Taylor instability over the timescale of the experiment.

I start by investigating the simplest topographic defect, a one-dimensional step of height $h_s$ that spans the full width of the substrate (see Fig.~\ref{fg:fig2}a inset, and SM~\cite{Note1}). Fig.~\ref{fg:fig2}a shows two chronophotographies of a pendant drop sliding under the same pre-wet surface with a step defect, at two slightly different inclination angles. Below a critical inclination angle $\alpha_c$, the pendant drop is stopped by the step. Above $\alpha_c$, the drop is slowed down but is able to pass (see also movie S1~\cite{Note1}). The step thus displays a pinning-like force that hinders the drop motion. Note that this pinning force is directional. In the experiment shown in Fig.~\ref{fg:fig2}a the drop climbs up the step. Inverting the step such that the drop descends it instead results in the drop accelerating over the step. 
In principle, this pinning force can depend on the droplet size $A$, the pre-wetting thickness $h_0$, the step height $h_s$ and its sharpness $\delta$ (see inset Fig.~\ref{fg:fig2}a). To explore this large parameter space efficiently, I turn to numerical simulations.

Taking advantage of the problem thinness, I use the lubrication approximation to describe the liquid film thickness evolution $h(x,y,t)$ over an uneven substrate with profile $s(x,y)$~\cite{Yiantsios:1989,Stillwagon:1990,Gaskell:2004,Lister:2010} inclined by an angle $\alpha$ (see Fig.~\ref{fg:fig1}b), but retain the non-linear full curvature $\kappa$ of the interface $h+s$~\cite{Wilson:1982,Lerisson:2020,Ledda:2021,Jambon:2021b}. The dimensionless thin-film equation in the Cartesian frame aligned with the substrate reads 
\begin{equation}
\begin{gathered}
\partial_{\bar{t}}\bar{h} + \widetilde{\alpha}\bar{h}^2 \partial_{\bar{x}}\bar{h} + (1/3)\mathbf{\bar{\nabla} \cdot} \left[\bar{h}^3\mathbf{\bar{\nabla}}\left(\bar{h}+\bar{s}\right) + \bar{h}^3 \mathbf{\bar{\nabla}}\bar{\kappa} \right]=0, \\
\bar{\kappa}=\mathbf{\bar{\nabla} \cdot}\left[\mathbf{\bar{\nabla}}\left(\bar{h}+\bar{s}\right)  \middle/\sqrt{1+\left(h_0\sqrt{\cos\alpha}/\ell_c \right)^2 \left(\mathbf{\bar{\nabla}}\left(\bar{h}+\bar{s}\right) \right)^2}\right],
\end{gathered}
\label{eq:lub_adim}
\end{equation}
after rescaling $\{x,y\}$ with $\ell_c/\sqrt{\cos\alpha}$, $\{h,s\}$ with $h_0$ the initial uniform thickness far from the drop and $t$ with $\tau=\eta\gamma/\left(h_0^3\rho^2g^2\cos^2\alpha\right)$. Here a bar indicates rescaled variables and the parameter $\widetilde{\alpha}=(\ell_c \tan\alpha)/(h_0\sqrt{\cos\alpha})$ accounts for the substrate inclination. I numerically solve Eq.~\eqref{eq:lub_adim} with COMSOL, using the initial condition $\bar{h}(\bar{x},\bar{y},0)=1+\bar{h_d}(\bar{x},\bar{y})$, where $h_d(x,y)$ is the profile of a static pendant drop (see SM~\cite{Note1} for numerical methods). 

\begin{figure}[t]
    \begin{center}
        \includegraphics[width=\columnwidth]{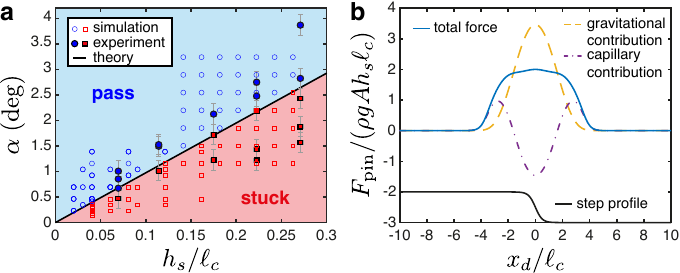}
        \caption{\textbf{a.} Phase diagram for the drop capture by a sharp step with $\delta<2\ell_c$ (see legend). 
        The black line and background colors are Eq.~\eqref{eq:critangle_step}. \textbf{b.} Dimensionless pinning force for a drop of size $A=\ell_c$ on a sharp step ($\delta=\ell_c$, black solid curve with abscissa $x/\ell_c$). The gravitational and capillary contributions to the total force are highlighted (see legend).
        } 
        \label{fg:fig3}
    \end{center}
\end{figure}

In Fig.~\ref{fg:fig2}b, I show a numerical reproduction of the experiment shown in panel a, where the step is idealized as an hyperbolic tangent: $s_{1d}(x,y)=(h_s/2) \left(1-\tanh(2x/\delta)\right)$ (see also movie~S2~\cite{Note1}). There is a critical inclination angle to pass the step $\alpha_c$ which matches the experimental one, despite the different step shape and sharpness. Varying the liquid parameters, i.e. the drop initial amplitude ($0.65 <A_0/\ell_c< 1.42$) and the pre-wetting film thickness ($0.04<h_0/\ell_c<0.08$), I observe an influence on the drop speed and growth rate, as previously shown on flat surfaces~\cite{Lister:2010,Jambon:2021b}. Yet, surprisingly they have no impact on the critical angle to pass the step $\alpha_c$. The topography, however, does impact the pinning process. For sharp enough steps $\delta \lesssim R$, with $R\approx 3.58\ell_c$ the drop radius~\cite{Jambon:2021b}, the critical angle appears independent of the sharpness, while for very smooth steps $\delta \gtrsim R$, it varies with it. The step height $h_s$ affect $\alpha_c$ in all cases: the higher the step, the larger the critical angle. In SM~\cite{Note1}, I show that the smooth step regime can be understood as a local slope variation such that $\alpha_c\sim h_s/\delta$. In the following, I focus on the more complex sharp regime and show in Fig.~\ref{fg:fig3}a a phase diagram for the drop behavior combining simulations where all parameters are varied but keeping $\delta<2\ell_c$ (points for different values of $\delta$, $h_0$, and $A_0$ are superimposed). I observe a linear relationship $\alpha_c\propto h_s$, which is confirmed by experiments done on five milled steps of different heights with $h_0\approx 60$ $\mu$m and $0.55 <A_0/\ell_c< 1.07$, also show in Fig.~\ref{fg:fig3}a.

Since there is no contact line, the origin of this pinning-like force has to be gravito-capillary. As the drop advances on the step, locally some of the drop volume is lifted up, which costs gravitational energy and can be seen as an energy barrier. In the sharp limit, the volume to be lifted over a distance $h_s$ is maximum when the apex of the drop reaches the step and can be estimated as $dV\sim 2 A R dx$. The energy cost is thus $dE_g\sim -\rho g (2 AR dx) h_s $ and the force is $F_{\mathrm{pin}}=-dE_g/dx \sim \rho g AR h_s$. Neglecting the droplet growth during the step crossing, the surface inclination generates a driving force $F_d=\rho g V \sin\alpha$ with $V\sim AR^2$. Equating the two yields for small angles $\alpha_c\sim h_s/R$. The radius of pendant drops being $R\approx 3.58\ell_c$~\cite{Jambon:2021b}, I recover the scaling observed in Fig.~\ref{fg:fig3}a. In this simple scaling argument, the change of liquid surface area due to the topography is neglected. Yet, it was shown to be determinant for microfluidic drops in channels with surface defects~\cite{Dangla:2011}. To go beyond, I have to compute the total energy barrier that also includes the capillary contribution.

In the reference frame aligned with the surface (see Fig.~\ref{fg:fig1}b), the gravitational energy has two contributions: one from the slope responsible for the driving force $F_d=\rho g V \sin\alpha$ and one from the surface unevenness $E_g=\iiint \rho g \cos(\alpha) z \;dx dy dz=\iint \frac{1}{2} \rho g\cos(\alpha)  \left((h+s)^2-s^2\right) \;dx dy$ (see SM~\cite{Note1}). The capillary energy is $E_c=\iint \gamma \sqrt{1+\left(\partial_x (h+s)\right)^2+\left(\partial_y (h+s)\right)^2} dx dy$. To compute the total energy $E=E_g+E_c$ as a function of the drop position $(x_d,y_d)$ with respect to the topography, I assume that the liquid thickness is $h(x,y)=h_0+h_d(x-x_d,y-y_d)$. Using the small slope approximation, and assuming a small inclination angle yields (see SM~\cite{Note1}) 
\begin{equation}
\begin{split}
\frac{E(x_d,y_d)}{\rho g}&=E_{g0}+ E_{c0} + \iint h_d(x-x_d,y-y_d) s(x,y) dx dy \\
&+ \ell_c^2 \iint \left(  \left. \partial_x h_d\right\rvert_{\substack{x-x_d\\y-y_d}} \partial_x s + \left. \partial_y h_d\right\rvert_{\substack{x-x_d\\y-y_d}} \partial_y s \right) dx dy.
\end{split}
\label{eq:energy}
\end{equation}
Here $E_{g0}$ and $E_{c0}$ are constant terms with respect to the drop position. The pre-wetting thickness $h_0$ enters in these terms and does not participate to the energy barrier. Rescaling $\{x,y\}$ with the capillary length $\ell_c$, $h_d$ with the drop amplitude $A$, and $s$ with the topography scale $h_s$ the pinning force is then 
\begin{equation}
\frac{\mathbf{F_{\mathrm{pin}}}(\hat{x_d},\hat{y_d})}{\rho g A h_s \ell_c}=-\hat{\nabla} \left[(\hat{h_d} * \hat{s}) + (\partial_{\hat{x}} \hat{h_d} * \partial_{\hat{x}} \hat{s}) + (\partial_{\hat{y}} \hat{h_d} * \partial_{\hat{y}} \hat{s}) \right].
\label{eq:force}
\end{equation}
Here $\hat{\cdot}$ denotes dimensionless variables, and $(*)$ the convolution operator $(f*g)(x_d,y_d)=\iint f(x-x_d,y-y_d) g(x,y) dxdy$. While not fully analytical, Eq.~\eqref{eq:force} can be integrated very efficiently by computing the 2d convolutions in Fourier space.

Performing this calculation for the steps $\hat{s}_{1d}$, I obtain the theoretical pinning force shown in Fig.~\ref{fg:fig3}b. The capillary contribution partially counteracts the gravitational one and the pinning force is maximum when the drop apex sits on the step. Extracting this maximum and equating it with the drop driving force $F_d\approx\rho g V \alpha$, I obtain the critical inclination angle for the drop to pass the step
\begin{equation}
\begin{split}
\alpha_c&= c_{1d} \frac{h_s}{\ell_c},\\
c_{1d}&=\frac{A\ell_c^2}{V}\max_{\hat{x_d},\hat{y_d}}\left(-\partial_{\hat{x}} \left[(\hat{h_d} * \hat{s}_{1d}) + (\partial_{\hat{x}} \hat{h_d} * \partial_{\hat{x}} \hat{s}_{1d}) \right]\right)\approx 0.1703.
\end{split}
\label{eq:critangle_step}
\end{equation}
The prefactor $c_{1d}$ is almost independent of the choice of static pendant drop $h_d$ to convolve due to their near self-similarity (see SM~\cite{Note1}). It also varies very slowly with $\delta$ in the sharp limit and I thus assume it to be constant in that regime. In Fig.~\ref{fg:fig3}a, I overlay the result of Eq~\eqref{eq:critangle_step} on numerical and experimental data and find an excellent agreement without any fitting parameter. I further test the model to see if it can capture the transition between sharp and smooth steps ($\delta> 2\ell_c$) in SM~\cite{Note1} and again find an very good agreement.

\begin{figure}[t]
    \begin{center}
        \includegraphics[width=\columnwidth]{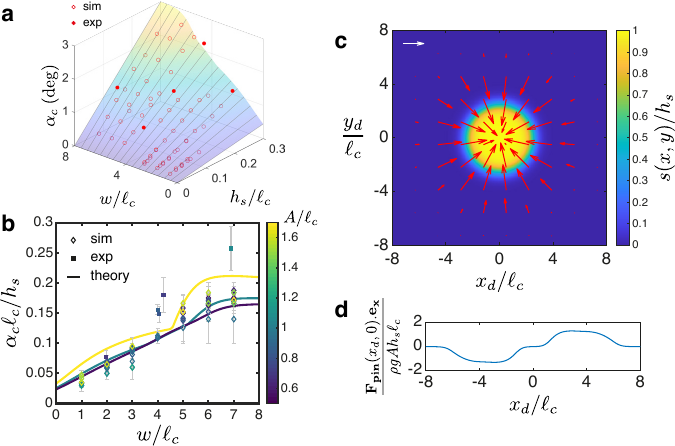}
        \caption{\textbf{a.} Critical angle $\alpha_c$ as a function of the dimensionless bump width $w/\ell_c$ and height $h_s/\ell_c$. The colored surface is Eq.~\eqref{eq:critangle_bump} for a pendant drop of size $A=\ell_c$. \textbf{b.} Reduced critical angle $\alpha_c \ell_c/h_s$ as a function of the dimensionless bump width $w/\ell_c$. Same data as a, the color represents the drop dimensionless height $A/\ell_c$ while on the bump. \textbf{c.} Dimensionless pinning force field $\mathbf{F_{\mathrm{pin}}}/(\rho g A h_s \ell_c)$ (arrows) for a bump with $w/\ell_c=4$ (background color with axes $\{x,y\}/\ell_c$) and a pendant drop of size $A=\ell_c$. The white arrow has a magnitude of $1$. \textbf{d.} Slice of the data in c through center of the defect.
        } 
        \label{fg:fig4}
    \end{center}
\end{figure}

I now extend the analysis to bidimensional steps, or bumps, $s_{2d}(x,y)=\frac{h_s}{2} \left(1-\tanh\left[\frac{2\sqrt{x^2+y^2}-w}{\delta}-1\right]\right)$. I focus on bumps with a size comparable to drops $w\lesssim 2R$ since very wide bumps $w\gg R$ will reduce to 1d steps. Performing simulations similar to the ones shown in Fig.~\ref{fg:fig2}b, I find qualitatively that bumps behave similarly to steps (see Movie~S3~\cite{Note1}). For small enough inclination angles, the drop is captured by the bump while for larger angles it is able to pass over it.  However, the capture here is omnidirectional, as expected from the symmetry of the defect. Turning the bump into a trough, on the opposite repels the drop (see Movie~S3~\cite{Note1}). 

Remaining in the sharp regime, I run simulations with $\delta=\ell_c$ to construct a multidimensional phase diagram from which I extract the critical angle $\alpha_c$ shown in Fig.~\ref{fg:fig4}a as a function of the bump height and width. As shown, wider and taller bumps generate a larger critical angle. The pre-wetting film thickness $h_0$ still does not affect the observed critical capture angle ($0.02<h_0/\ell_c<0.06$) while the drop sizes $A$ now slightly impacts it. Equation~\eqref{eq:force} predicts a linear increase with $h_s$. I thus plot in Fig.~\ref{fg:fig4}b the critical angle rescaled by the bump height $\alpha_c \ell_c/h_s$ as a function of the dimensionless bump width  $w/\ell_c$ and observe a good collapse of the numerical data. I compare these numerical results to experiments with five milled surfaces, also shown in Fig.~\ref{fg:fig4}a-b. The experimental critical angles are in line with numerics, although systematically slightly higher likely due to the finite time used to determine whether the drop was pinned, and the different bump shape (see SM~\cite{Note1}).

To understand the width dependence, I use Eq.~\eqref{eq:force}, as in the step case, but now with $\hat{s}_{2d}(\hat{x},\hat{y})$. I show in Fig.~\ref{fg:fig4}c-d the theoretical force field for a pendant drop of size $A=\ell_c$ on a bump of width $w/\ell_c=4$. The bump defect acts as an omnidirectional attractor for the drop with a range $\sim w/2+R$. The maximum pinning force does not occur at the center of the bump but on an annulus, in line with the observed off-centered drop pinning (see movie~S3~\cite{Note1}). Equating the maximum pinning force in the slope direction $x$ with the driving force, I calculate the critical angle $\alpha_c = c_{2d}(w/\ell_c,h_d) \frac{h_s}{\ell_c}$ with 
\begin{equation}
\begin{split}
c_{2d}(w/\ell_c,h_d) & =\frac{A\ell_c^2}{V}\max_{\hat{x_d},\hat{y_d}}\left(-\partial_{\hat{x}} \left[(\hat{h_d} * \hat{s}_{2d}) + \right.\right. \\
&\left.\left. (\partial_{\hat{x}} \hat{h_d} * \partial_{\hat{x}} \hat{s}_{2d})  + (\partial_{\hat{y}} \hat{h_d} * \partial_{\hat{y}} \hat{s}_{2d}) \right]\right).
\end{split}
\label{eq:critangle_bump}
\end{equation}
I compare Eq.~\eqref{eq:critangle_bump} to simulations and experiments in Fig.~\ref{fg:fig4}a-b and find a good agreement. The width function $c_{2d}(w/\ell_c,h_d)$ increases with the bump width until it eventually plateau around the step value $c_{1d}$ for bumps larger than the drop $w\gtrsim 2R$. The influence of the pendant drop size $A/\ell_c$, while still small is more pronounced than in the step case, especially as the drop gets larger (see Fig.~\ref{fg:fig4}b and SM~\cite{Note1}). This size dependence is not so clear in simulation data, perhaps because the drop profile shifts from its flat static case assumed in the model.

\begin{figure}[t]
    \begin{center}
        \includegraphics[width=\columnwidth]{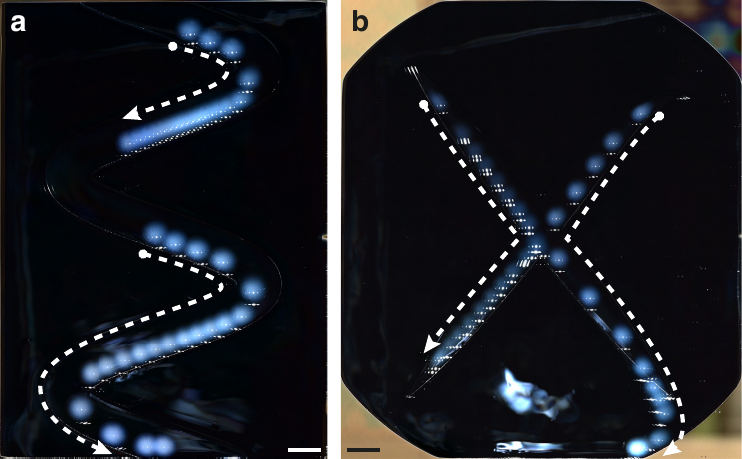}
        \caption{Bottom view chronophotography of two pendant droplets sliding under a pre-wet tilted plate milled with an elevated path ($\alpha\approx 1$ deg from top to bottom, $h_0\approx 44$ $\mu$m): sinusoidal in \textbf{a} and X shaped in \textbf{b} (see Movie~S4~\cite{Note1}). Each droplet's trajectory is indicated by a white dashed arrow. White dye was used for visualization. Scale bars $1$ cm, time interval $15$ min. 
        } 
        \label{fg:fig5}
    \end{center}
\end{figure}

Now that I understand the gravito-capillary pinning force generated by elementary defects, I can design surfaces to guide pendant droplets along a chosen path. One can either use trenches that repel the drop, or elevate the surface along the desired path to attract the droplet~\cite{Abbyad:2011}. The guiding element(s) width $w$ and height $h_s$ can be tuned using Fig.~\ref{fg:fig4}b or Eq.~\eqref{eq:force} to provide sufficient pinning for the desired inclination angle. As a proof of concept, I mill a surface with an elevated sinusoidal path, of width $w=12$ mm and height $h_s=250$ $\mu$m. As shown in Fig.~\ref{fg:fig5}a and movie~S4~\cite{Note1}, pendant droplets under this pre-wet surface inclined by $\alpha\approx 1$ deg follow the prescribed sinusoidal path. In Fig.~\ref{fg:fig5}b I show a more complex X shaped path that features a branching point ($h_s=250$ $\mu$m, varying width, $\alpha\approx 1$ deg). Two drops traveling on the converging branches of the X will merge if they meet at the branching point, but will continue on their respective branch if they pass sequentially (see movie~S4~\cite{Note1}).

In summary, I have demonstrated that topographic defects impact the motion of pendant droplets on the underside of wet substrates, despite their lack of contact line. Using numerical simulations, experiments, and theory, I have shown that defects generate a gravito-capillary pinning force capable of capturing a drop; for the drop to slide over the defect, some liquid has to change altitude and the liquid surface has to deform. I propose a model for the pinning force, Eq.~\eqref{eq:force}, that predicts the critical capture angle observed in experiments and numerical simulations quantitatively without any fitting parameter. Finally, I show how to use these findings to design topographies that control the motion of pendant drops.

These results not only shed light on an unusual form of droplet pinning but could also be harnessed to design surfaces that control the position and motion of pendant drops~\cite{Marthelot:2018,Jambon:2021a}. Besides, gravito-capillary pinning is relevant to thin film dynamics on natural non-smooth surfaces and could affect the formation of some geomorphological patterns like curtains and draperies in natural caves~\cite{Ledda:2021}.

\begin{acknowledgments}
I thank T. Perrin and C. Hasson from the Drahi-X Novation Center for their help in milling the surfaces. I am grateful to P.G. Ledda, F. Gallaire, and P.-T. Brun for their feedback on the manuscript.
\end{acknowledgments}

\end{document}


\makeatletter
\let\ps@titlepage\ps@plain
\makeatother

\title{Supplemental Material for: \\
Gravito-capillary pinning of pendant droplets under wet uneven surfaces}
\author{Etienne Jambon-Puillet}
\affiliation{LadHyX, CNRS, Ecole Polytechnique, Institut Polytechnique de Paris, Palaiseau, France}

\date{\today}

\begin{abstract}
In this supplementary document, I provide experimental, numerical, and theoretical methods, as well as extra numerical and theoretical results.
\end{abstract}

\maketitle

\section{Methods}
\begin{figure*}[t]
    \begin{center}
        \includegraphics[width=0.86\textwidth]{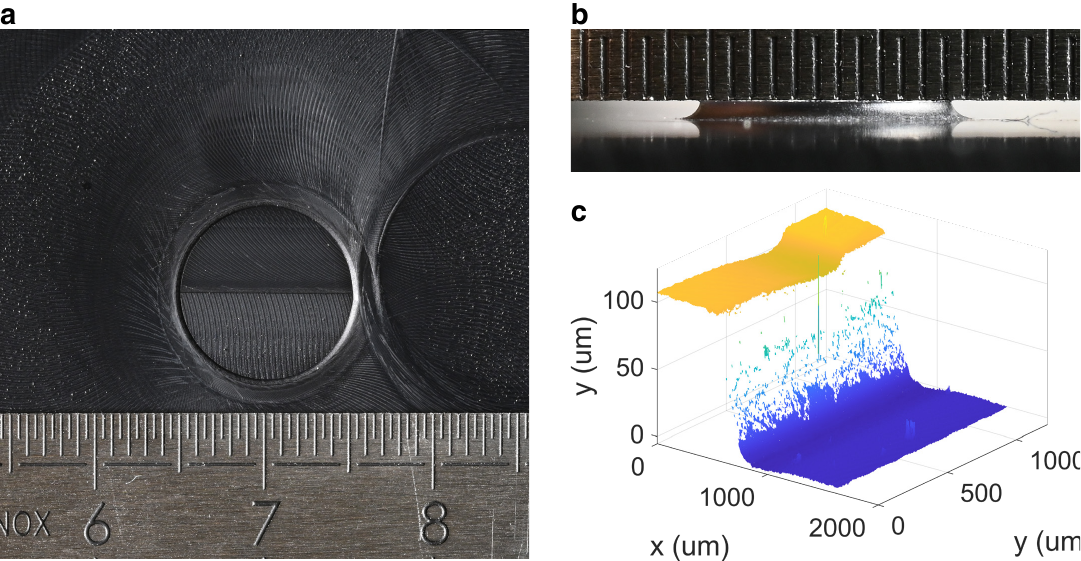}
        \caption{\textbf{a.} Top view of a surface with a milled bump of width $w=10$ mm and height $h_s=250$ $\mu$m. Small and large toolmarks due to the milling process are visible. \textbf{b.} Side view of a milled bump of width $w=6$ mm and height $h_s=400$ $\mu$m highlighting the bump edge. The ruler smallest graduation is $0.5$ mm. \textbf{c.} Profilometry reconstruction of a step of height $h_s=100$ $\mu$m (outliers have been filtered). A toolmark is visible on the top surface.
        } 
        \label{fg:fig_exp}
    \end{center}
\end{figure*}

\noindent\textbf{Experiments.} Textured surfaces were milled out of black PMMA with a Datron M8Cube, then analyzed with a Zygo Newview 7100 profilometer, and photographed. Fig.~\ref{fg:fig_exp}a-b show top and side view pictures of representative bump defects, and Fig.~\ref{fg:fig_exp}c show the 3d profile of a 100 $\mu$m step defect. While all defects are designed perfectly sharp, their edge has a finite width that results from the milling process. The edge profile is quite different from the hyperbolic tangent used in the simulations and has a width $\delta$ of the order of the defect height. On flat regions of the surface, periodic toolmarks are visible. The largest ones are $5-10$ $\mu$m in depth and set the uncertainty for $h_s$.

Silicone oil v1000 (density $\rho=971$ kg/m$^3$, surface tension $\gamma=20.5$ mN/m, viscosity $\eta=1$ Pa.s) was used in the experiments. To prepare the uniform coating $h_0$, the edges of the milled surfaces were first covered with tape over a width of $\sim 1$ cm. Oil was then spin-coated (Polos SPIN150i) over the surface and the tape subsequently pealed to remove the edge bead that spontaneously form during spin-coating. The surface was then weighted with a precision scale to determine the film thickness $h_0=m/(\rho S)$ with $m$ the film mass, and $S$ the surface area not covered by the tape (measured with a camera). While defects are known to impact spin-coating~\cite{Stillwagon:1990,Gaskell:2004}, perturbations are usually localized and the initial thickness is thus assumed uniform. A single spin-coating procedure was used for all the steps and bumps experiments which resulted in a thickness $h_0\approx 62\pm 4$ $\mu$m. A small amount of white oil dye was used on path experiments for visualization and a smaller thickness $h_0\approx 44\pm 3$ $\mu$m was selected to avoid dripping.

The coated surface is then attached upside-down with screws to an arm mounted on a rotation stage. A weighted fishing line was attached to the arm to calibrate the vertical direction. Using a syringe, a drop of oil was injected on the surface to form a pendant drop. The dispensed volume was difficult to control as some liquid dripped along the needle, but the drop amplitude was systematically measured and varied in the range $0.7 <A< 1.6$ mm. For surfaces with a step defect, the drop was deposited far from the step and the arm was rotated to the desired angle $\alpha$ (measured optically). The drop dynamic was then recorded from the side for about an hour with a camera (see movie S1). For surfaces with a bump defect, the drop was deposited directly on the bump and recorded from the side while the surface was tilted incrementally. A waiting period of at least $10$ min between angle increments was used, and the angle was increased until the drop detached from the defect. The measurement uncertainty on the angle $\alpha$ is estimated to be around $0.15$ deg. 

All image analysis was performed with ImageJ. To construct the experimental images and supplementary movies, the experimental setup vibrations are filtered by registering the stacks with rigid transformations using the plugin ``Linear Stack Alignment with SIFT'' or ``Descriptor-based series registration''~\cite{Preibisch:2010}. The chronophotographies are made with standard deviation projections.\\

\begin{figure*}[t]
    \begin{center}
        \includegraphics[width=\textwidth]{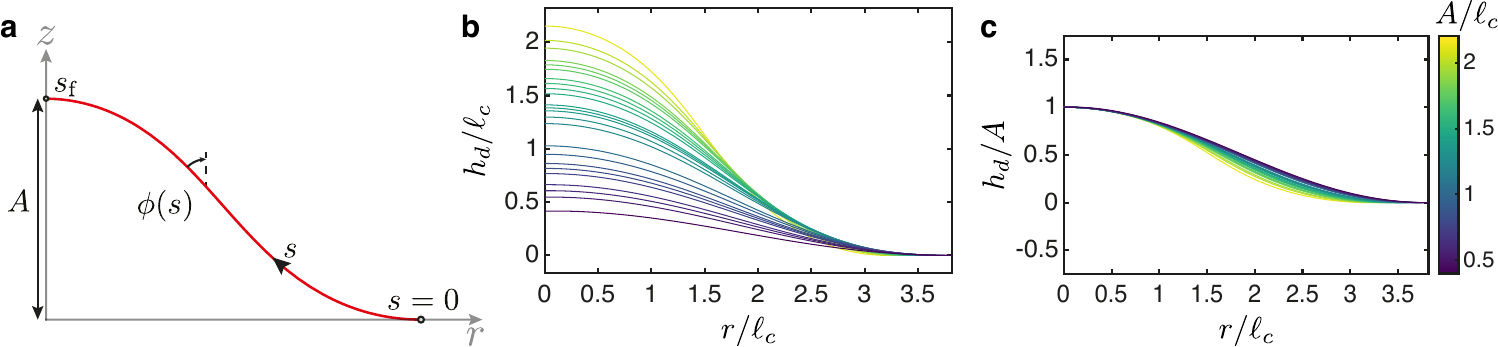}
        \caption{\textbf{a.} Schematic of a static pendant drop defining the coordinates used
for the numerical integration of Eq.~\eqref{eq:YoungLap}. \textbf{b.} Static pendant drops profiles. The color codes the drop amplitude (see legend in c). \textbf{c.} Same profiles as b, but rescaled by the drop size $A$.
        } 
        \label{fg:fig_YoungLap}
    \end{center}
\end{figure*}

\noindent\textbf{Simulations.} Finite element simulations were performed with the commercial software Comsol v6.1 and v6.2. The simulations solve the dimensional version of Eq.~(1) using the silicone oil values for an easier comparison with experiments:
\begin{equation}
\begin{gathered}
\frac{\partial h}{\partial t} + \mathbf{\nabla} \cdot \left[\frac{\rho g h^3}{3\eta}\left(\cos\alpha\mathbf{\nabla}\left(h+s\right) + \ell_c^2 \mathbf{\nabla}\kappa + \sin\alpha \mathbf{e_x}\right) \right]=0, \\
\kappa=\mathbf{\nabla} \cdot\left[\frac{\mathbf{\nabla}\left(h+s\right)} {\sqrt{1+ \left(\frac{\partial \left(h+s\right)}{\partial x }\right)^2 + \left(\frac{\partial \left(h+s\right)}{\partial y }\right)^2}}\right].
\end{gathered}
\label{eq:lub_dim}
\end{equation}
Note that Eq.~\eqref{eq:lub_dim} assumes small slopes, also for the substrate topography $s(x,y)$. Therefore, the step and bump defects cannot be as sharp in the numerics as they are in the experiments where the small residual sharpness comes from the milling process. While the step equation $s_{1d}$ uses a standard hyperbolic tangent, the bump equation $s_{2d}(x,y)=\frac{h_s}{2} \left(1-\tanh\left[\frac{2\sqrt{x^2+y^2}-w}{\delta}-1\right]\right)$ is written such that the bump width $w$ is defined around the top of the bump rather than at mid-height ($s_{2d}(w/2,0)\approx 0.88 h_s$). This choice allows to compare the numerical results with milled experimental surface where the width is defined at the top in the milling process.

The static pendant drop profile $h_d(x,y)$ supplied as initial condition $h(x,y,0)=h_0+h_d(x,y)$ to the solver was computed numerically from the Young-Laplace equation with the appropriate boundary conditions~\cite{Marthelot:2018} using Mathematica (shooting method): 
\begin{equation}
\begin{gathered}
\frac{\mathrm{d}^2 \phi\left(s\right)}{\mathrm{d} s^2}=\frac{-\cos\phi\left(s\right)}{\ell_c^2}+\frac{\mathrm{d}}{\mathrm{d} s}\left[\frac{\cos\phi\left(s\right)}{r(s)}\right]\\
\frac{\mathrm{d} h_d(s)}{\mathrm{d} s}=\cos\phi\left(s\right),\quad \frac{\mathrm{d} r(s)}{\mathrm{d} s}=\sin\phi\left(s\right),\\
\quad h_d(0)=0, \quad \phi(0)=-\pi/2,\\
r(s_\mathrm{f})=0, \quad \phi(s_\mathrm{f})=-\pi/2, \quad h_d(s_\mathrm{f})=A.
\label{eq:YoungLap}
\end{gathered}
\end{equation}
Here, \{$r(s)$, $h_d(s)$\} are the (cylindrical) coordinates of the drop surface, $\phi(s)$ is the local angle that the tangent makes with the vertical and $s$ is the arc-length as defined in Fig.~\ref{fg:fig_YoungLap}a. The value of $s_\mathrm{f}$ is a priori unknown and is determined by the additional boundary condition. The drops shapes (see Fig.~\ref{fg:fig_YoungLap}b) are then imported in the FEM solver and used for the initial condition.

For all quantitative steps and bumps simulations, I take advantage of the system symmetry to only solve half of the domain shown in movies S2 and S3. I use rectangular domains with no-flux boundary conditions, a square mesh with quadratic Lagrange elements for $h$ and for $\kappa$ (resolved separately) of size $0.1\ell_c$ or smaller. The simulation domain has dimensions \{$L_x\approx 53.3\ell_c$, $L_y\approx 17.77\ell_c$\} for steps and \{$L_x\approx 35.54\ell_c$, $L_y\approx 17.77\ell_c$\} for bumps. Simulations are run until at least $t=10\tau$ with two stop conditions: the simulation is stopped if the drop reaches the size that would make it drip (if $\max_{x,y}(h)>2.2\ell_c$) or if it gets too close to the boundary of the simulation domain (distance of the drop apex to the boundary smaller than $8\ell_c$). In total $329$ simulations are performed for steps and $786$ for bumps. Of those, only the ones done in the sharp regime are shown in the main text. A few, additional simulations were performed for supplementary movies and designing the paths to guide the drop. Those are done with different domain sizes, sometimes at a slightly lower resolution (typically $0.15\ell_c$), and for the trough and paths where there is no longer a symmetry plane, Eq.~\eqref{eq:lub_dim} was solved on the full domain. \\

\begin{figure}[t]
    \begin{center}
        \includegraphics[width=1\columnwidth]{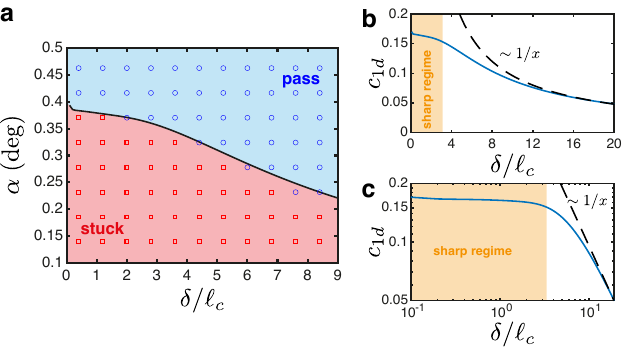}
        \caption{\textbf{a.} Numerical phase diagram for the drop capture by a step of fixed height $h_s=0.04\ell_c$ and varying width $\delta/\ell_c$ (simulations are done with $h_0=0.04\ell_c$, and $A_0\approx\ell_c$). Red squares indicate drops that remained stuck on the step, blue circles drops that passed the step, the black line and background colors are the theory, i.e. Eq.~(4), for the same pendant drop \textbf{b.} Theoretical rescaled critical angle as function of the step sharpness (blue solid curve). The sharp regime where $c_{1d}$ is almost independent of $\delta$ is highlighted, the dashed black curve is a guide to the eye for the smooth step limit. \textbf{c.} Same data plotted in log-log.
        } 
        \label{fg:fig_delta}
    \end{center}   
\end{figure}

\noindent\textbf{Theory.} In a Cartesian frame $(x',y',z')$ aligned with gravity, the gravitational energy is $E_g^{tot}=\iiint \rho g z' \;dx' dy' dz'$. Rotating the frame by an angle $\alpha$ to align it with the surface (see Fig.~1b) gives the new frame $(x,y,z)$. The Jacobian determinant of this rotation is 1 such that the gravitational energy now reads $E_g^{tot}=\iiint \rho g (z\cos\alpha-x\sin\alpha) \;dx dy dz$. The second term $E_g^*=-\iiint \rho g\sin(\alpha) x \;dx dy dz$ gives a force 
\[\mathbf{F_g^*}=-\mathbf{\nabla} E_g^*=\rho g\sin(\alpha) \partial_x \iiint  x \;dx dy dz \mathbf{e_x}\]
which turns out to be the usual gravitational driving force on slopes $F_d=\rho V g \sin\alpha$. The first term $E_g=\iiint \rho g \cos(\alpha) z \;dx dy dz$ is the one described in the main text. Integrating once between $z=s$ and $z=h+s$, I get $E_g=\frac{1}{2}\rho g \cos\alpha\iint \left((h+s)^2-s^2\right) dx dy$. Now introducing $h(x,y)=h_0+h_d(x-x_d,y-y_d)$ and expanding yields
\[\frac{E_g}{\rho g \cos\alpha}= \frac{1}{2}\iint \left(h_0^2+h_d^2+2 h_0 s+ 2 h_0 h_d + 2 h_d s\right) dx dy.\]
All the terms but the last one are constant with respect to the drop position ($x_d$,$y_d$) and I define 
\[E_{g0}=\frac{1}{2}\iint \left(h_0^2+h_d^2+2 h_0 s+ 2 h_0 h_d \right) dx dy.\]
The capillary energy is $E_c=\gamma S$ with $S=\iint \sqrt{1+\left(\partial_x (h+s)\right)^2+\left(\partial_y (h+s)\right)^2} dx dy$ the liquid surface area. I assume small slopes for $h$ and $s$ such that 
\[S\approx\iint \left[1+\frac{1}{2}\left(\left(\partial_x (h+s)\right)^2+\left(\partial_y (h+s)\right)^2\right)\right] dx dy.\]
Now introducing $h(x,y)=h_0+h_d(x-x_d,y-y_d)$ and expanding again yields
\begin{equation*}
\begin{split}
S&\approx\iint \left[ 1+\frac{1}{2}\left(\left(\partial_x h_d\right)^2+\left(\partial_x s\right)^2+ \left(2 \partial_x h_d \partial_x s\right)\right.\right. \\
&\left.\left. + \left(\partial_y h_d\right)^2+\left(\partial_y s\right)^2+ \left(2 \partial_y h_d \partial_y s\right)\right) \right] dx dy.
\end{split}
\end{equation*}
Here also, only the cross terms vary with respect to the drop position ($x_d$,$y_d$) and I define 
\[E_{c0}=\ell_c^2\iint \left[1+\frac{1}{2}\left(\left(\partial_x h_d\right)^2+\left(\partial_x s\right)^2+\left(\partial_y h_d\right)^2+\left(\partial_y s\right)^2\right)\right]dx dy.\]
I thus arrive to Eq.~(2) assuming a small tilt, i.e. $\cos\alpha\approx 1$.

To compute the pinning force and rescaled critical angles $c_{1d}$ and $c_{2d}$, I first generate the surface $\hat{s} (\hat{x},\hat{y})$ on a Cartesian grid with a resolution of at least $0.05\delta$. I then select a pendant drop profile (computed from Eq.~\eqref{eq:YoungLap}, see Fig.~\ref{fg:fig_YoungLap}b), rescale it by its maximum, as the model requires (see Fig.~\ref{fg:fig_YoungLap}c), and then project it on a Cartesian grid of the same resolution as the surface $\hat{s}$. I finally compute the convolutions using the fft method for speed. The computations are done on MATLAB R2021b and I use the toolbox ``FFT-based convolution'' from Bruno Long.

\section{Additional numerical and theoretical results}
\begin{figure}[t]
    \begin{center}
        \includegraphics[width=0.86\columnwidth]{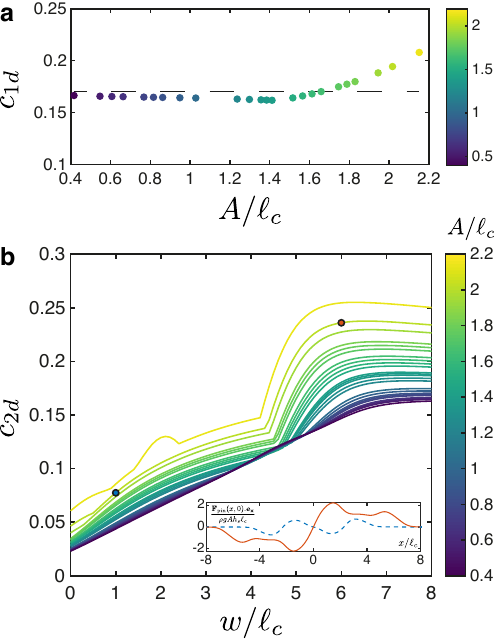}
        \caption{\textbf{a.} Value of the prefactor $c_{1d}$ for a sharp step with $\delta=\ell_c$ (see Eq.~(4)) as a function of the pendant drop chosen. The dashed line indicates the average value used in the main text $c_{1d}=0.1703$. \textbf{b.} Value of the function $c_{2d}(w/\ell_c, h_d)$ for sharp bumps with $\delta=\ell_c$ (see Eq.~(5)). The different pendant drops are color coded with colorbars identical to Fig.~\ref{fg:fig_YoungLap}b-c. Inset: Slice of the dimensionless pinning force field passing trough the center of the bump for the two points highlighted in the main figure: drop of size $A\approx 2\ell_c$ and the bumps of width $w=1\ell_c$ (blue) and $w=6\ell_c$ (orange).       
        } 
        \label{fg:fig_A}
    \end{center}
\end{figure}

\noindent\textbf{Step sharpness.} In the main text, I focus on the sharp regime when $\delta\lesssim R$. Fig.~\ref{fg:fig_delta}a shows a numerical phase diagram for steps of constant height $h_s$ but varying sharpness $\delta$. For $\delta<2\ell_c$, no influence is visible with the angle increment chosen to construct the phase diagram. For larger values of $\delta$, the critical angle eventually decreases. I plot the result of the model, i.e. Eq.~(4), on Fig.~\ref{fg:fig_delta}a and find here also an excellent agreement. The model can thus capture the transition to the smooth step regime as well. In Fig.~\ref{fg:fig_delta}b-c, the model is pushed to larger values of $\delta/\ell_c$ to evidence the smooth step regime where $\alpha_c\sim h_s/\delta$. The two regimes are particularly visible in log-scale (see Fig.~\ref{fg:fig_delta}c).\\

\noindent\textbf{Influence of the drop choice in the model.} To theoretically compute $c_{1d}$ or $c_{2d}$, one can use any possible pendant drop shape $h_d (x,y)$. However, as shown in Fig.~\ref{fg:fig_YoungLap}c, static pendant drops are almost self-similar and therefore the dimensionless profiles $\hat{h_d} (\hat{x_d},\hat{y_d})$ used in the convolution are not so different. Moreover, given that the volume of pendant drops varies as $V\approx 0.88 A R^2$ and that $R\approx 3.58\ell_c$ \cite{Jambon:2021b}, this yield a model where the drop size influence on the pinning force is of second order. In Fig.~\ref{fg:fig_A}a, I show the computed values of $c_{1d}$ for the pendant drops shown in Fig.~\ref{fg:fig_YoungLap}b-c. The prefactor remains in a narrow range except for very large drops on the verge of dripping ($A\gtrsim 2\ell_c$). 

For bumps, as shown in Fig.~\ref{fg:fig_A}b, the function $c_{2d}(w/\ell_c,h_d)$ is slightly more impacted by the drop choice, especially for large ones here as well. Notably, the shape of the curve varies significantly with the appearance of a kink around $w=5\ell_c$. This kink is a signature of the appearance of a second maximum in in the pinning force field close to the bump center, that eventually overtakes the initial one closer to the bump edge as shown in Inset of Fig.~\ref{fg:fig_A}b.

%